\documentclass[a4paper,11pt]{article}
\usepackage{pos}
\usepackage{bm}
\usepackage{orcidlink}
\usepackage{stackengine}
\usepackage{graphicx}
\usepackage{subcaption}
\usepackage{tikz-feynman}
\usepackage{pgfplots}
\usepgflibrary{plotmarks}
\title{Double $\phi$-meson production via $\bar{p}p$ scattering}
\author[a]{Dayoung Lee}
\author*[a]{Seung-il Nam}
\author[b]{Jung Keun Ahn}
\affiliation[a]{Department of Physics, Pukyong National University (PKNU), Busan 48513, Korea}
\affiliation[b]{Department of Physics, Korea University, Seoul 02841, Korea}
\emailAdd{sinam@pknu.ac.kr}
\emailAdd{ldyoung0421@pukyong.ac.kr}
\emailAdd{ahnjk@korea.ac.kr}
\abstract{We investigate double $\phi$ production in $\bar{p}p$ reactions near threshold using an effective Lagrangian approach, highlighting a significant OZI rule violation through hadronic processes. Our model includes $t$- and $u$-channel contributions from the nucleon and $s$-wave $N^*$ resonances ($N^*(1535,1650,1895)$), as well as the pentaquark-like state $P_s(2071)$. In the $s$-channel, scalar ($f_0$), tensor ($f_2$), and pseudoscalar ($\eta(2225)$) mesons are incorporated. The $N$, $N^*$, $f_0$, and $f_2$ contributions enhance the cross-section and generate peak structures near $W=2.2$ GeV, consistent with JETSET data, while cusp structures arise from $\bar{\Lambda}\Lambda$ and $\bar{\Sigma}\Sigma$ threshold openings. Polarization observables, such as the spin density matrix element (SDME), further elucidate the underlying hadronic mechanisms.}
\FullConference{The 21st International Conference on Hadron Spectroscopy and Structure (HADRON2025)\\
 27 - 31 March, 2025\\
Osaka University, Japan\\}
\begin{document}
\maketitle
\section{Introduction}
In constituent quark models, the proton and antiproton contain only up and down (anti)quarks, while the $\phi$ meson is almost a pure $\bar{s}s$ state. The reaction $\bar{p}p \to \phi\phi$ is therefore OZI-suppressed, as it requires disconnected quark lines and gluonic transitions. A conventional explanation involves intermediate $\omega\omega$ states and $\phi-\omega$ mixing, predicting a cross-section around 10 nb. However, JETSET experimental data shows a much larger cross-section of $(2-4) \mu$b, indicating a significant OZI violation~\cite{JETSET:1994evm,JETSET:1994fjp,JETSET:1998akg}. Several theoretical explanations have been proposed: contributions from glueballs or tetraquark states with $\bar{s}s$ content, instanton-induced interactions, and hadronic rescattering processes involving intermediate meson or baryon channels~\cite{Ke:2018evd,Lu:2019ira}. Glueball candidates predicted by lattice QCD and QCD sum rules fall within the accessible energy range, though their contribution remains uncertain~\cite{Chen:2005mg,Narison:1996fm,Chen:2021bck}. Additionally, strange sea-quark components in the nucleon wavefunction may allow connected diagrams via shake-out mechanisms without violating the OZI rule, potentially yielding cross-sections on the order of hundreds of nb~\cite{Ellis:1994ww}. Hadronic rescattering processes, such as $\bar{p}p \to \bar{K}K \to \phi\phi$ or $\eta\eta \to \phi\phi$, are also OZI-allowed and may enhance the observed rate~\cite{Lu:1992xd}. Within a hadronic model, the effective coupling $g_{\phi NN}$ is sizable, attributed to coupled-channel unitarization involving $K^*\Lambda$, $\omega N$, etc~\cite{Khemchandani:2011et}. Recent studies suggest that $t$-channel $N^*(1535)$ exchange and $s$-channel scalar/tensor mesons ($f_0$, $f_2$) can account for structures near $W \approx 2.2$ GeV in the cross-section~\cite{Xie:2014tra,Xie:2007qt}. 

This work applies a tree-level effective Lagrangian approach, including $N^*(1535,1650,1895)$ and the pentaquark-like $P_s(2071)$ in $t$- and $u$-channels, and scalar/tensor mesons ($f_0$, $f_2$) and $\eta(2225)$ in the $s$-channel. The model reproduces threshold enhancement and resonance peaks in the JETSET data. Cusp structures at $W \approx 2M_\Lambda$ and $2M_\Sigma$ arise from $\bar{\Lambda}\Lambda$ and $\bar{\Sigma}\Sigma$ channel openings. Contributions from $\eta$ and $P_s$ are negligible. Furthermore, spin density matrix elements (SDMEs) are computed to analyze the polarization of the final-state $\phi$ mesons, offering additional sensitivity to the reaction mechanism. These observables help constrain resonance couplings and validate the model through fits to both unpolarized and polarized data. The paper provides a foundation for future spin-dependent amplitude analyses to probe OZI violation in this channel further.
\section{Theoretical Framework}
\begin{figure}[h]
\begin{center}
\includegraphics[width= 11cm]{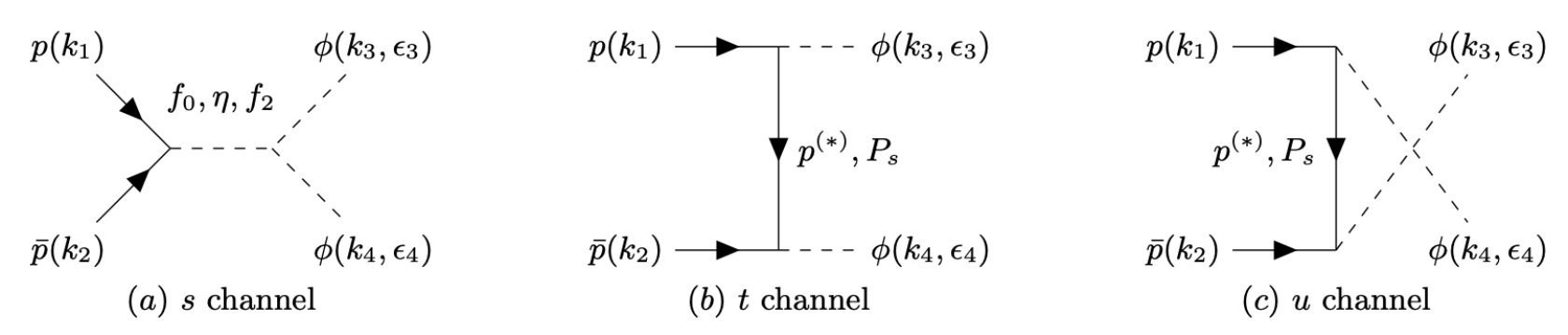}\hspace{0.4cm}
\includegraphics[width= 3.2cm]{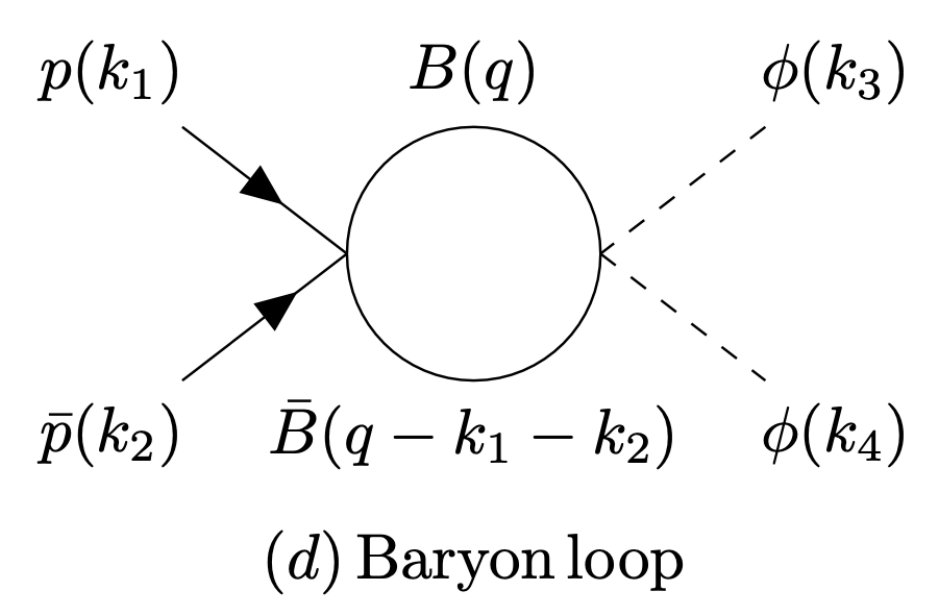}
\caption{The relevant Feynman diagrams illustrate the $(s,t,u)$-channel amplitudes for $\bar{p}p\to\phi\phi$ shown in panels (a), (b), and (c), respectively. Solid lines represent (anti)proton and its resonances in the diagrams, while dashed lines represent scalar and tensor mesons. The four momenta ($k_i$) and polarizations ($\epsilon_i$) for the particles are also defined. The last one (d) denotes the baryon-loop contribution for $\bar{B}B$ channel opening for $B=(\Lambda,\Sigma)$.}
\label{FIG0}
\end{center}
\end{figure}
We study the $\bar{p}p \to \phi\phi$ reaction using an effective Lagrangian approach, focusing on $s$-, $t$-, and $u$-channel Feynman diagrams in Fig.~\ref{FIG0}. The interaction Lagrangians involve scalar, pseudoscalar, vector, and tensor mesons, as well as nucleons and spin-$3/2$ resonances. For instance:
\begin{eqnarray}
\mathcal L_{S NN} &=& g_{S NN}\bar{N}SN+\text{h.c.}, \,\,\,
\mathcal L_{SVV} = \frac{g_{S \phi\phi}}{m_\phi}F_{\mu\nu}F^{\mu\nu}S, 
\cr
\mathcal L_{T NN} &=& -\frac{ig_{TNN}}{M_N}\bar{N}(\gamma_\mu\partial_\nu + \gamma_\nu\partial_\mu)N T^{\mu\nu}, 
\cr
\mathcal L_{VNN} &=& -g_{VNN}\bar{N}\left[\gamma_\mu - \frac{\kappa_{VNN}}{2M_N} \sigma_{\mu\nu}\partial^\nu \right]\Gamma_5 N V^\mu.
\end{eqnarray}
We employ the phenomenological form factors to consider the spatial extension of the hadrons:
\begin{equation}
F^h_x = \frac{\Lambda_h^4}{\Lambda_h^4 + (x - M_h^2)^2}, \quad x = s, t, u.
\end{equation}

To account for intermediate baryon loops like $\bar{B}B$ shown in Fig.~\ref{FIG0}, we include one-loop diagrams with point-like vertices:
\begin{equation}
\mathcal{L}_{4B} = \frac{g_{4B}}{M_N^2} \bar{B}\bar{B}BB, \quad
\mathcal{L}_{VBVB} = \frac{g_{VBVB}}{M_N^3} \bar{B}F_{\mu\nu}F^{\mu\nu}B.
\end{equation}

Polarization observables such as spin density matrix elements (SDMEs) are used to study helicity structures:
\begin{eqnarray}
\rho^0_{\lambda\lambda'} &=& \frac{1}{2N_T} \sum_{\lambda_{\bar{p}},\lambda_p,\lambda_{\phi_4}}
\mathcal{M}_{\lambda{\bar{p}}\lambda_p\lambda\lambda_{\phi_4}}
\mathcal{M}^{\lambda{\bar{p}}\lambda_p\lambda'\lambda_{\phi_4}}, 
\cr
\rho^4_{\lambda\lambda'} &=& \frac{1}{N_L} \sum_{\lambda_{\bar{p}},\lambda_p}
\mathcal{M}_{\lambda{\bar{p}}\lambda_p\lambda 0}
\mathcal{M}^{\lambda{\bar{p}}\lambda_p\lambda' 0}.
\end{eqnarray}

These SDMEs are frame-dependent (helicity, Adair, GJ) and allow comparison with experimental angular distributions, enabling a deeper understanding of the hadronic production mechanism of double $\phi$ mesons. More detailed explanations for the theoretical framework can be found in Ref.~\cite{Lee:2024opb}.
\section{Numerical results and Discussions}
In this section, we present the numerical results and discuss the key findings. For the $s$-channel, we include the scalar and tensor mesons $f_0(2020, 2100, 2200)$ and $f_2(1950, 2010, 2150)$, along with the pseudoscalar $\eta(2225)$, which is known to couple strongly to $\phi\phi$~\cite{ParticleDataGroup:2022pth}. The reduced couplings $g_\Phi = g_{\Phi\phi\phi}g_{\Phi NN}$ are listed in Table~\ref{TAB1}, with values adopted due to limited experimental constraints.
\begin{table}[b]
\begin{center}
\tiny
\begin{tabular}{ |c||c|c|c|c|c|c|} 
\hline
&$f_0(2020)$&$f_0(2100)$&$f_0(2200)$&$f_2(1950)$&$f_2(2010)$&$f_2(2150)$\\
\hline
 $M-i\Gamma/2$ [MeV]&$1982-i218$&$2095-i143.5$&$2187-i103.5$&$1936-i232$&$2011-i101$&$2157-i76$\\
 \hline
 $g_{(S,P,T)}$ & \multicolumn{3}{c|}{$0.115$} & \multicolumn{3}{c|}{$-0.1$}\\
\hline
\end{tabular}
\end{center}
\caption{Relevant meson coupling constants for the $s$-channel contributions.}
\label{TAB1}
\end{table}

For the $t$- and $u$-channels, we include nucleon resonances such as $N^*(1535,1650,1895)$ and a possible pentaquark $P_s(2071)$. Their couplings, including complex values for $g_{\phi NN^*}$, are listed in Table~\ref{TAB2}. The tensor interaction $\kappa_{N^*}$ is set to zero due to a lack of data, except for $\kappa_{\phi NN}=-1.65$. A common cutoff $\Lambda_{h,\mathrm{loop}}=(550,300)$ MeV is used in form factors.
\begin{table}[b]
\begin{center}
\tiny
\begin{tabular}{|c||c|c|c|c|c|c|c|} 
\hline
&$N$&$N^*(1535)$& \multicolumn{2}{c|}{$N^*(1650)$} & \multicolumn{2}{c|}{$N^*(1895)$}&$P_s(2071)$\\
\hline
$M-i\Gamma/2$ [MeV]&$938-i0$&$1504-i55$&$1668-i28$&$1673-i67$&$1801-i96$&$1912-i54$&$2071-i7$\\
\hline
$g_{\phi NN^{(*)}}$&$-1.47$&$1.4+i2.2$&$4.1-i2.7$&$4.5+i5.2$&$2.1+i1.8$&$0.9-i0.2$&$0.14+i0.2$\\
\hline
 \end{tabular}
 \end{center}
\caption{Relevant nucleon coupling constants for the $t$- and $u$-channel contributions.}
\label{TAB2}
\end{table}

\begin{figure}[h]
\begin{tabular}{cc}
\topinset{(a)}{\includegraphics[width= 3.5cm]{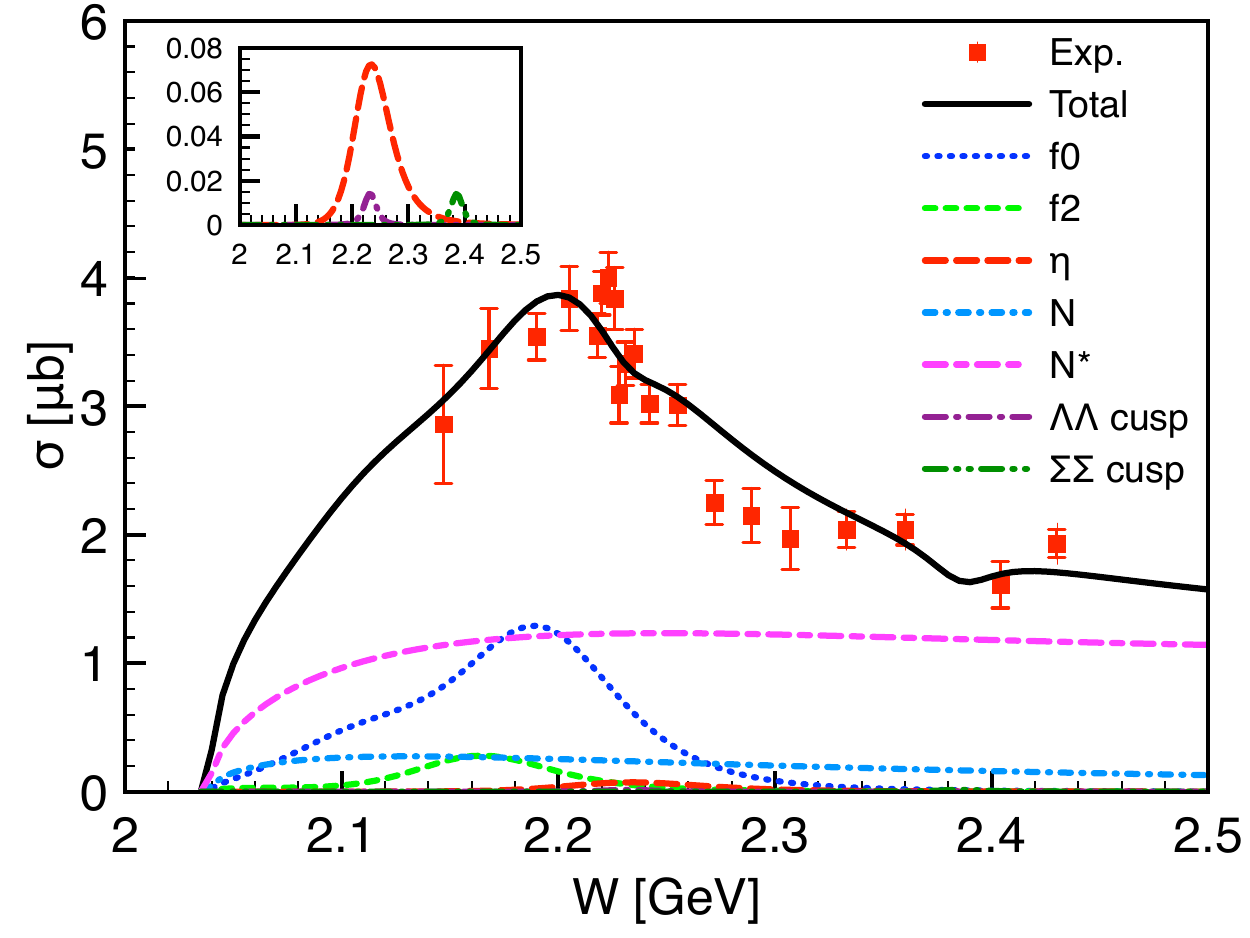}}{-0.2cm}{0.5cm}
\topinset{(b)}{\includegraphics[width= 3.5cm]{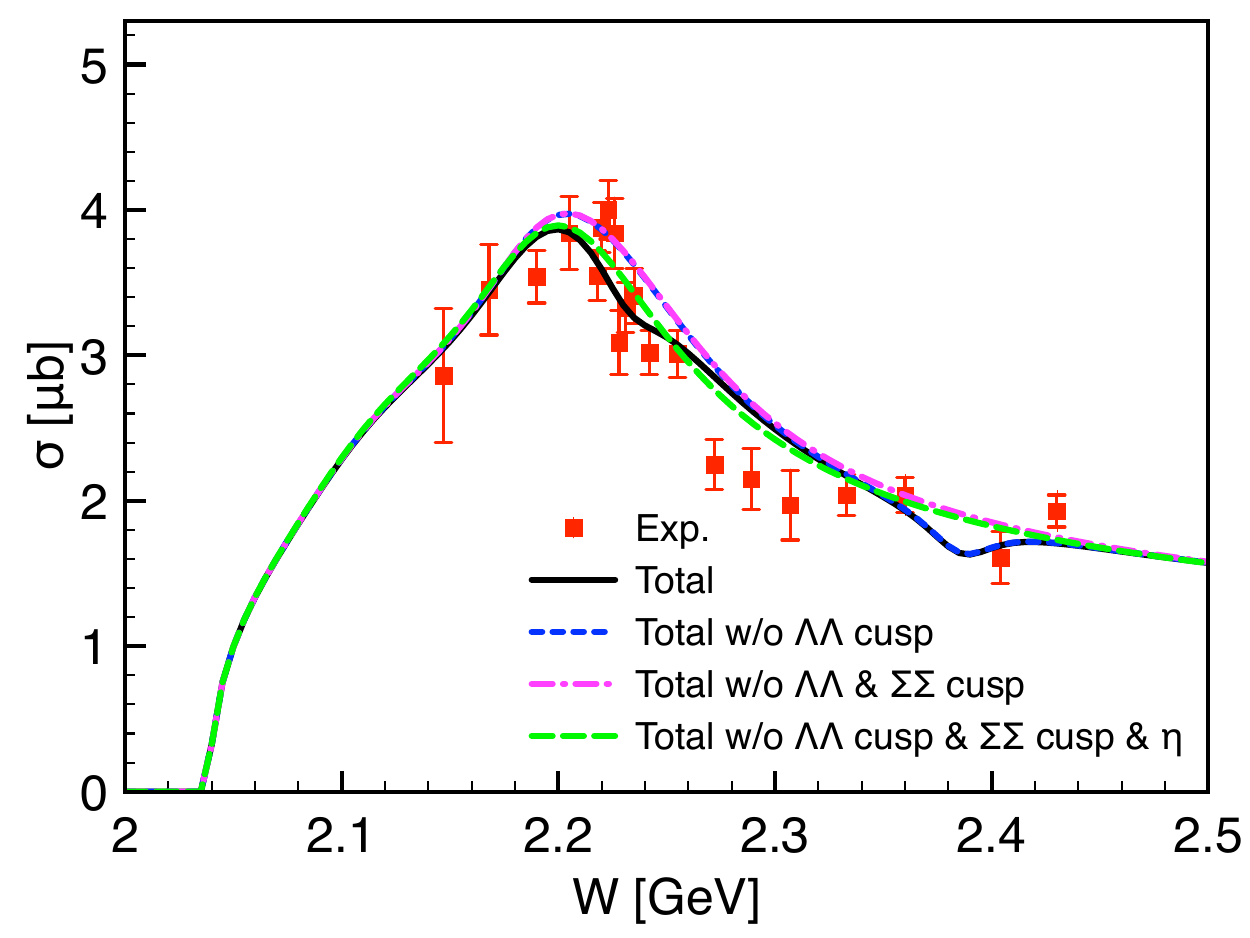}}{-0.2cm}{0.5cm}
\topinset{(c)}{\includegraphics[width= 3.5cm]{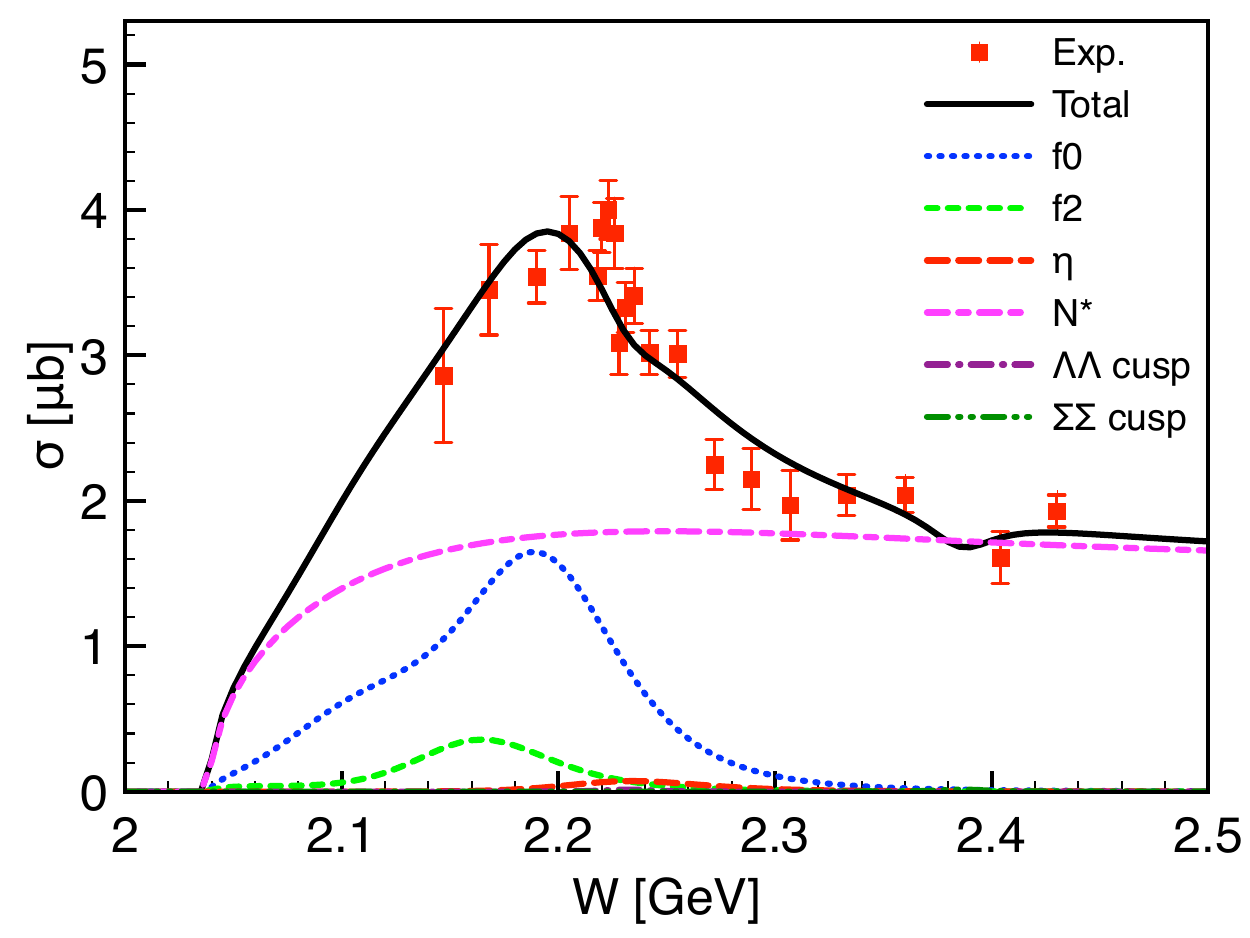}}{-0.2cm}{0.5cm}\,\,\,\,
\topinset{(d)}{\includegraphics[width= 3.5cm]{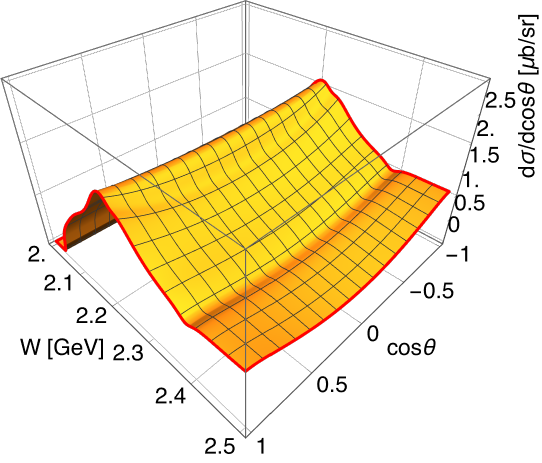}}{-0.2cm}{0.5cm}
\end{tabular}
\caption{(Color online) (a) Total cross-sections $\sigma\equiv\sigma_{\bar{p}p\to\phi\phi}$ as functions of $W$, showing each contribution separately. Experimental data are taken from Ref.~\cite{JETSET:1994evm,JETSET:1994fjp,JETSET:1998akg}. (b) Those with and without the cusp effect in addition to the $\eta$ contribution. (c) Those without the ground-state nucleon ($N$) contribution. (d) Angular-dependent differential cross-section $d\sigma/d\cos\theta$ as a function of $W$ and $\cos\theta$.}
\label{FIG2}
\end{figure}
Pannel (a) of Fig.~\ref{FIG2} shows the total cross-section $\sigma_{\bar{p}p\to\phi\phi}$ as a function of $W$. Near threshold, the $N$ contribution dominates, while $N^*$ resonances grow with energy. The $f_0$ and $f_2$ mesons create a bump around $W=2.2$ GeV. The cusp structures near $2M_\Lambda$ and $2M_\Sigma$ arise from $\bar{\Lambda}\Lambda$ and $\bar{\Sigma}\Sigma$ loops, as confirmed in panel (b). We also test the $\eta$ effect, which alone cannot explain the observed structure. Panel (c) tests excluding the $N$ contribution, which removes the threshold shoulder and shifts the shape above $W=2.4$ GeV. Panel (d) presents the angular distribution $d\sigma/d\cos\theta$, showing symmetry about $\theta = \pi/2$ due to identical final-state mesons.

\begin{figure}[h]
\begin{center}
\begin{tabular}{ccc}
\topinset{Adair $\rho^{\lambda}_{00}$}{\includegraphics[width= 2.8cm]{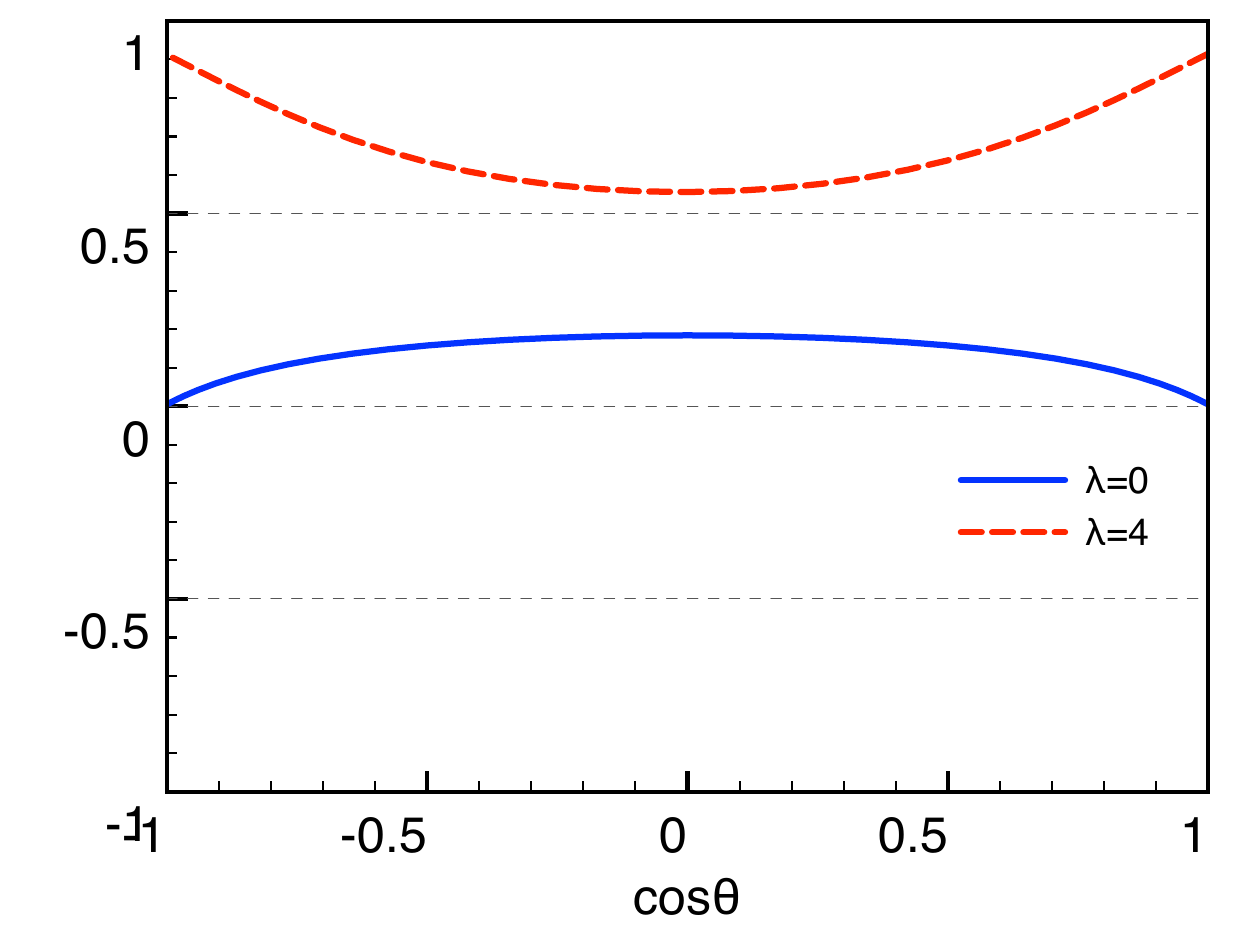}}{-0.4cm}{0.5cm}
\topinset{Adair $\rho^{\lambda}_{10}$}{\includegraphics[width= 2.8cm]{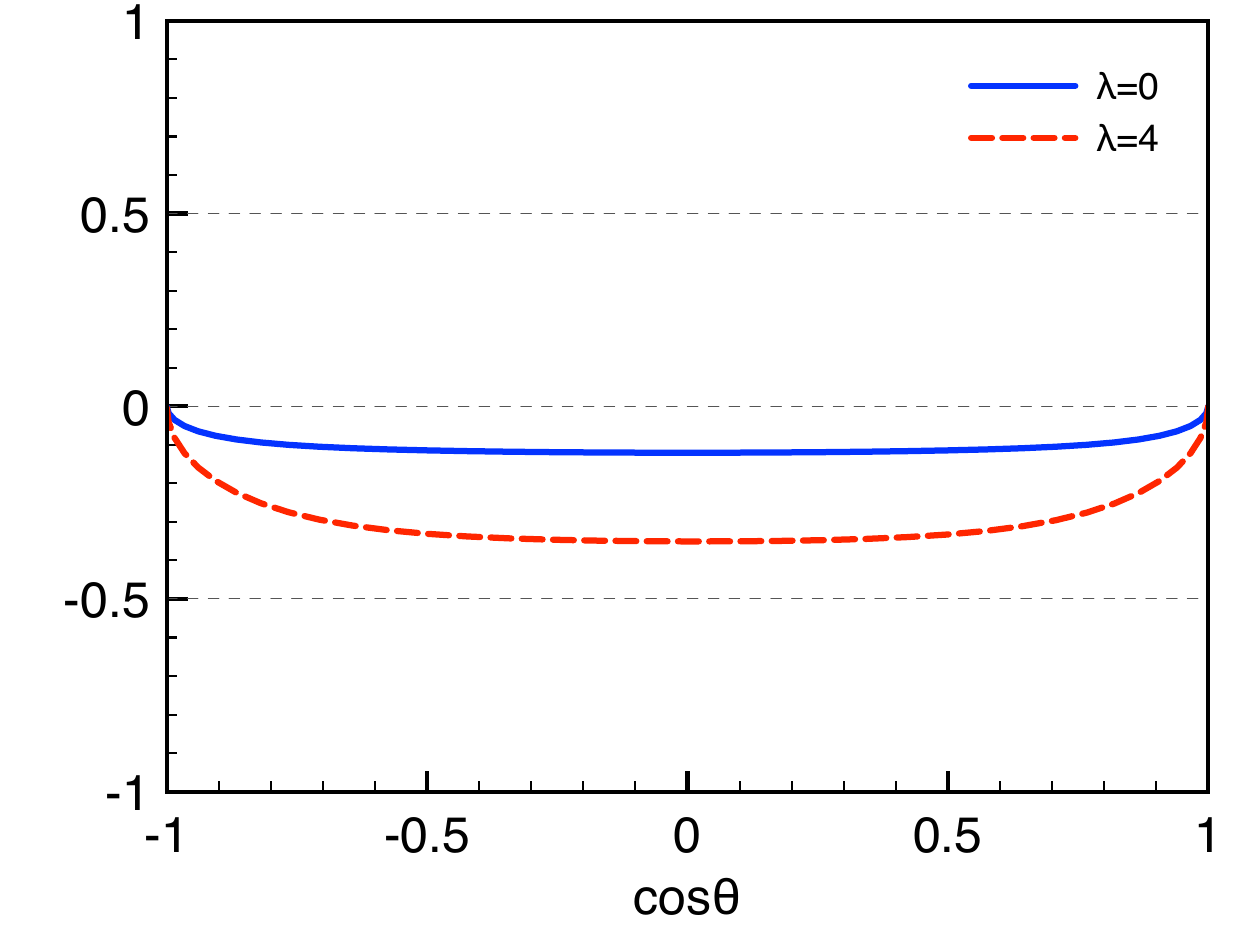}}{-0.4cm}{0.5cm}
\topinset{Adair $\rho^{\lambda}_{1-1}$}{\includegraphics[width= 2.8cm]{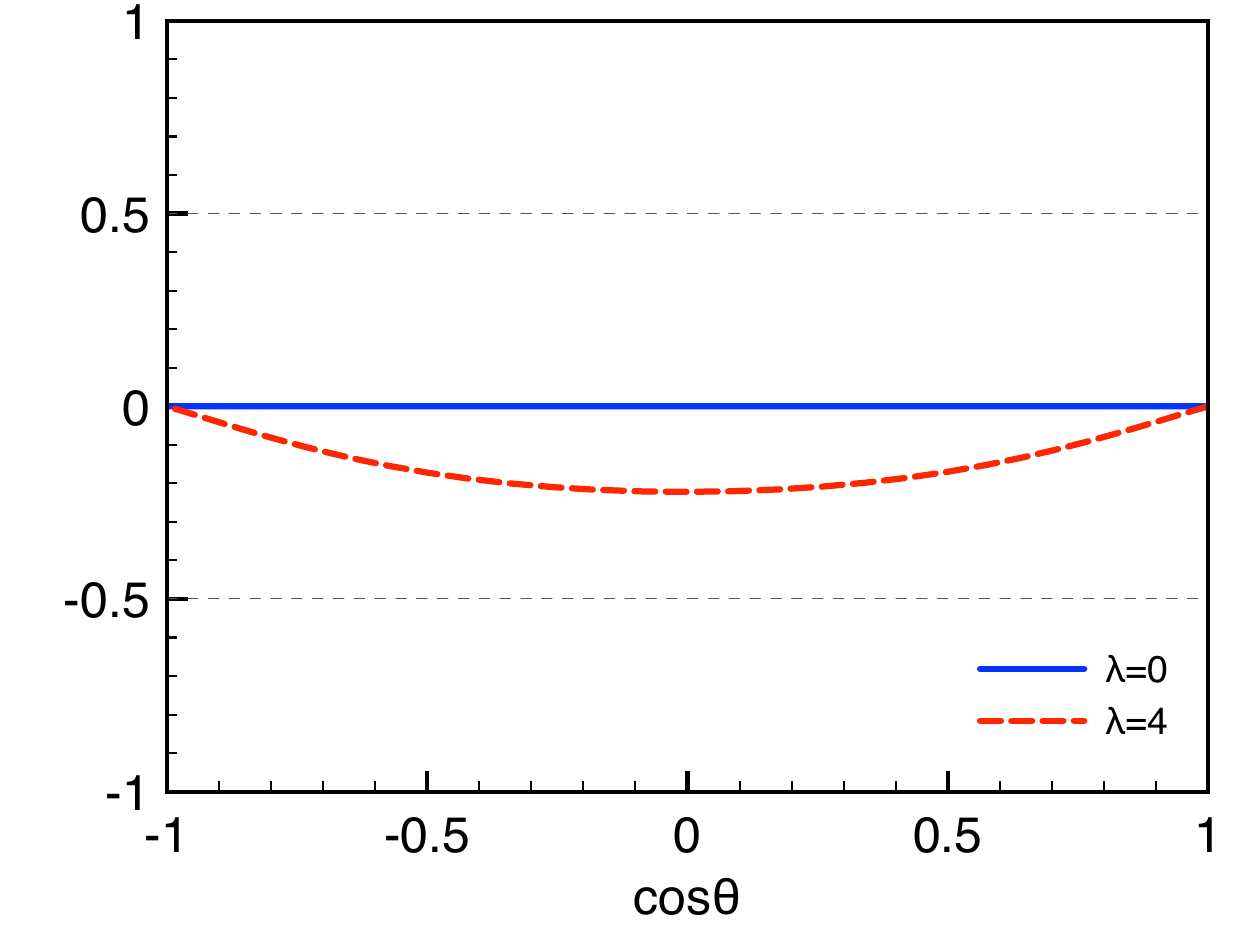}}{-0.4cm}{0.5cm}
\topinset{Helicity $\rho^{\lambda}_{00}$}{\includegraphics[width= 2.8cm]{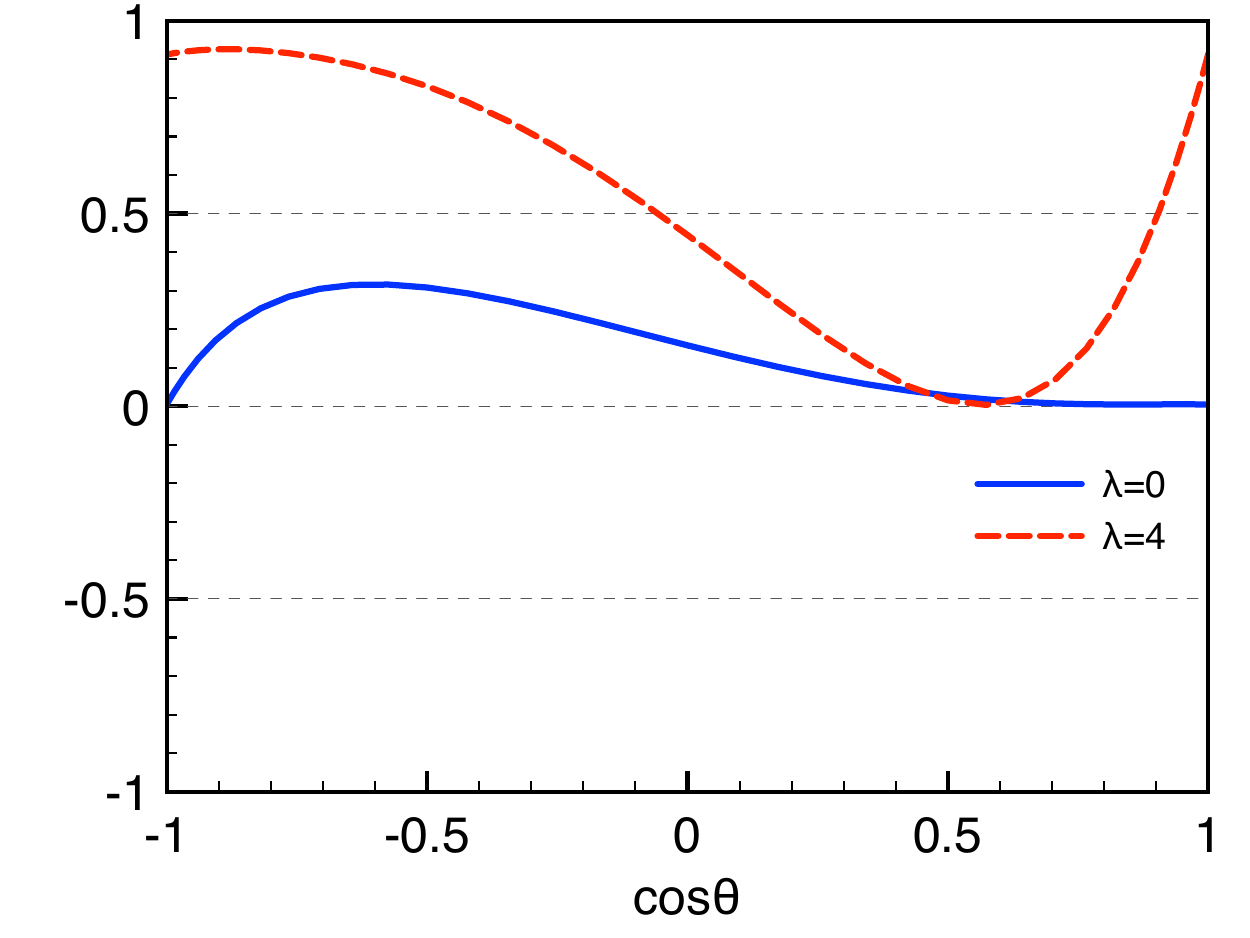}}{-0.4cm}{0.5cm}
\topinset{Helicity $\rho^{\lambda}_{10}$}{\includegraphics[width= 2.8cm]{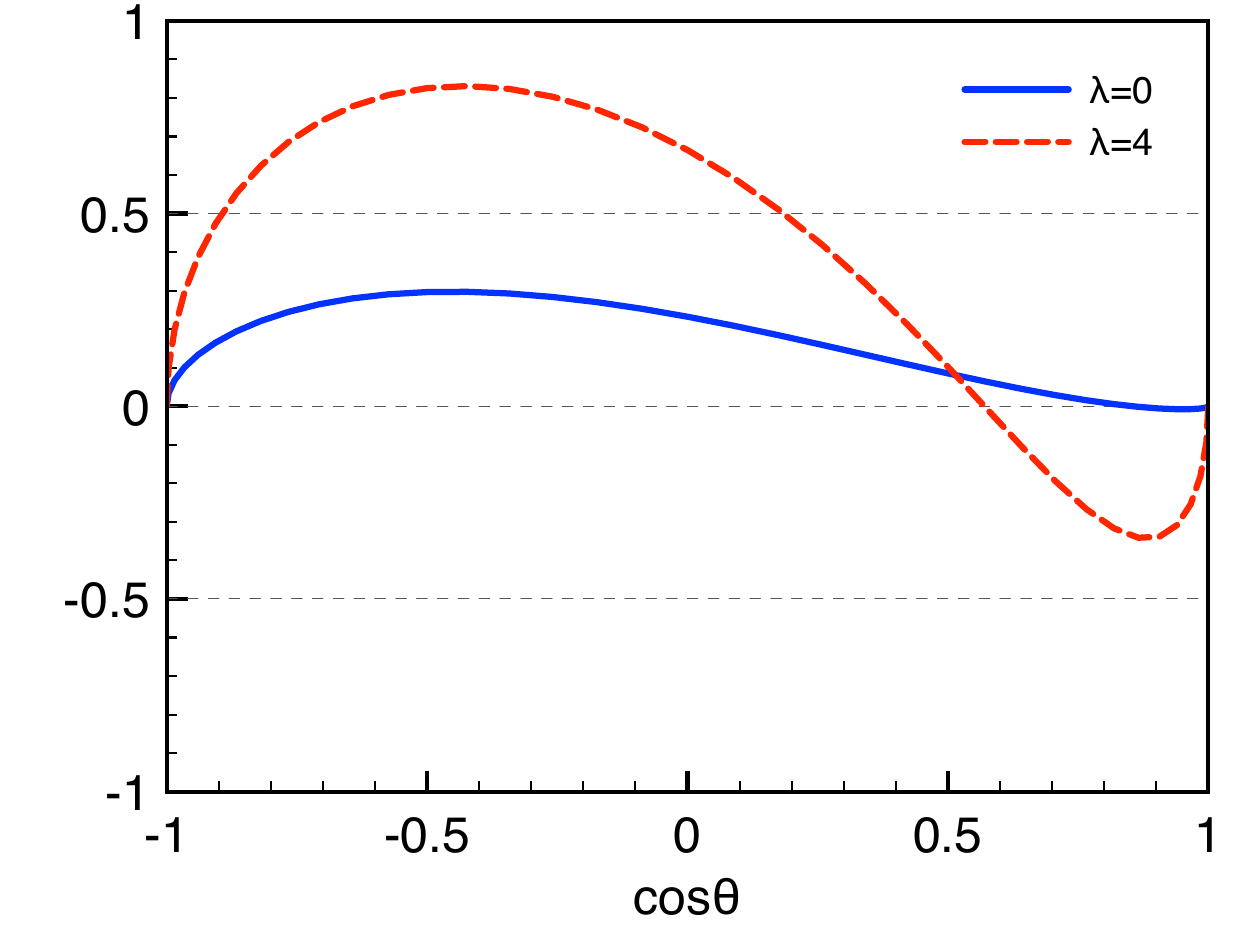}}{-0.4cm}{0.5cm}
\end{tabular}
\begin{tabular}{ccc}
\topinset{Helicity $\rho^{\lambda}_{1-1}$}{\includegraphics[width= 2.8cm]{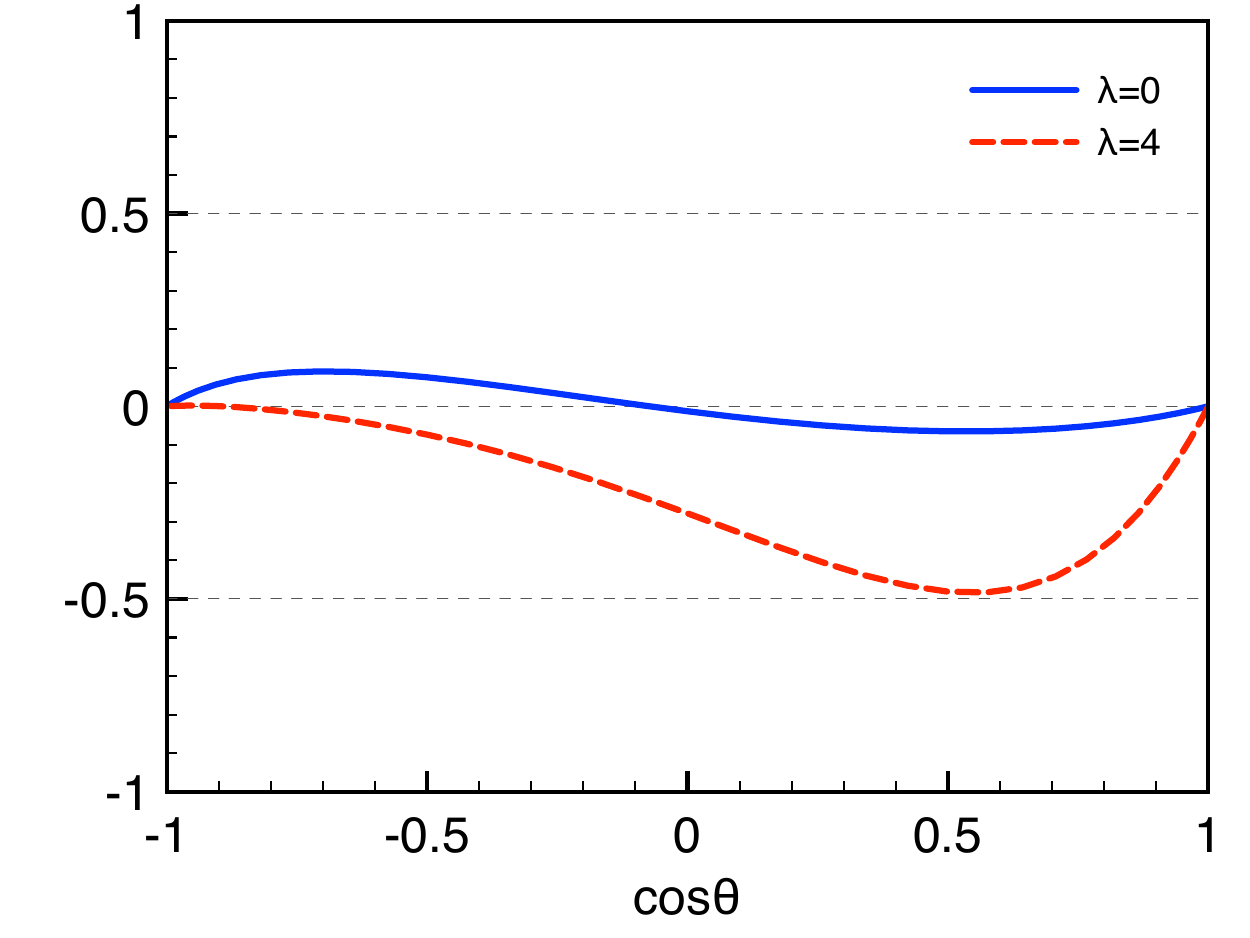}}{-0.4cm}{0.5cm}
\topinset{GJ $\rho^{\lambda}_{00}$}{\includegraphics[width= 2.8cm]{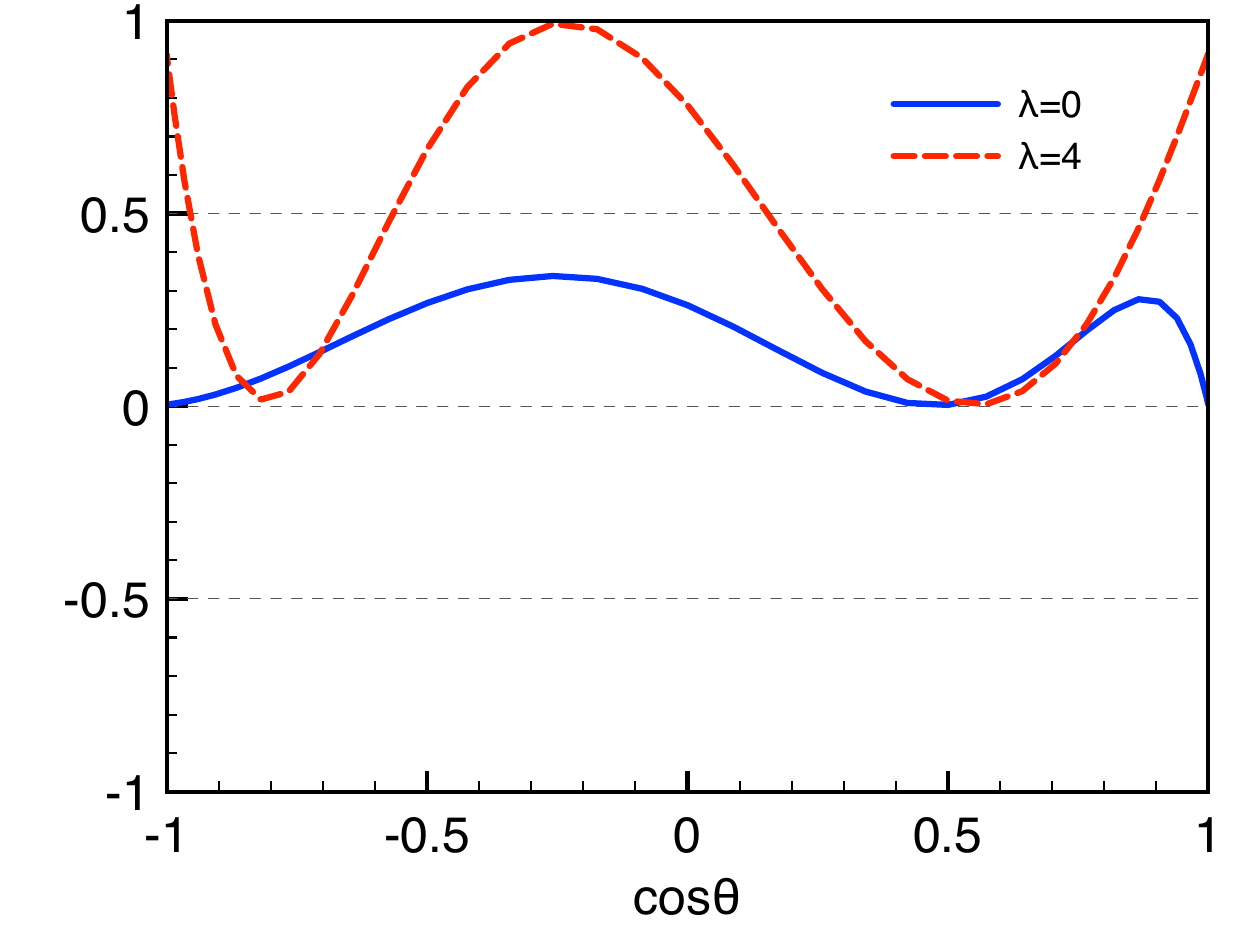}}{-0.4cm}{0.5cm}
\topinset{GJ $\rho^{\lambda}_{10}$}{\includegraphics[width= 2.8cm]{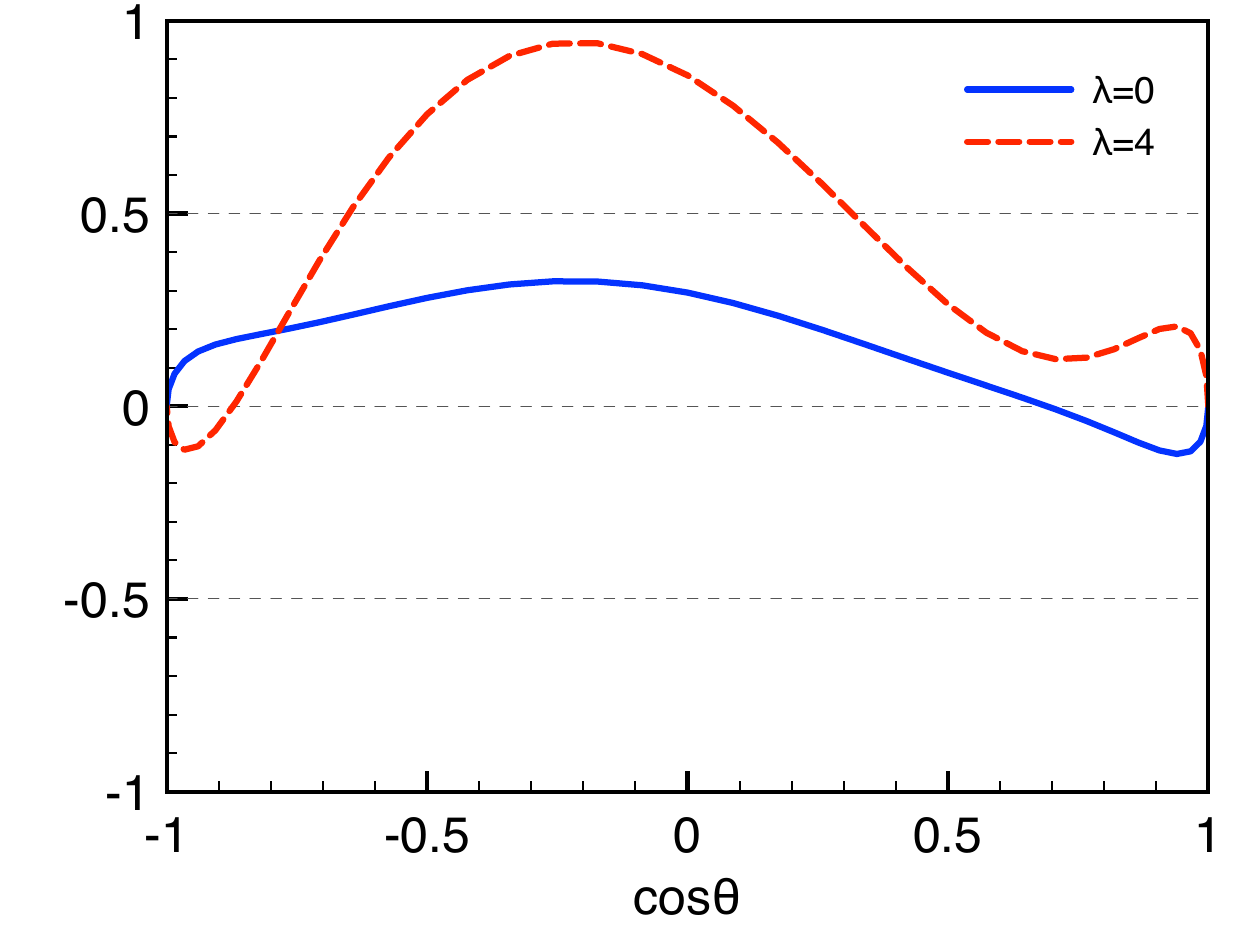}}{-0.4cm}{0.5cm}
\topinset{GJ $\rho^{\lambda}_{1-1}$}{\includegraphics[width= 2.8cm]{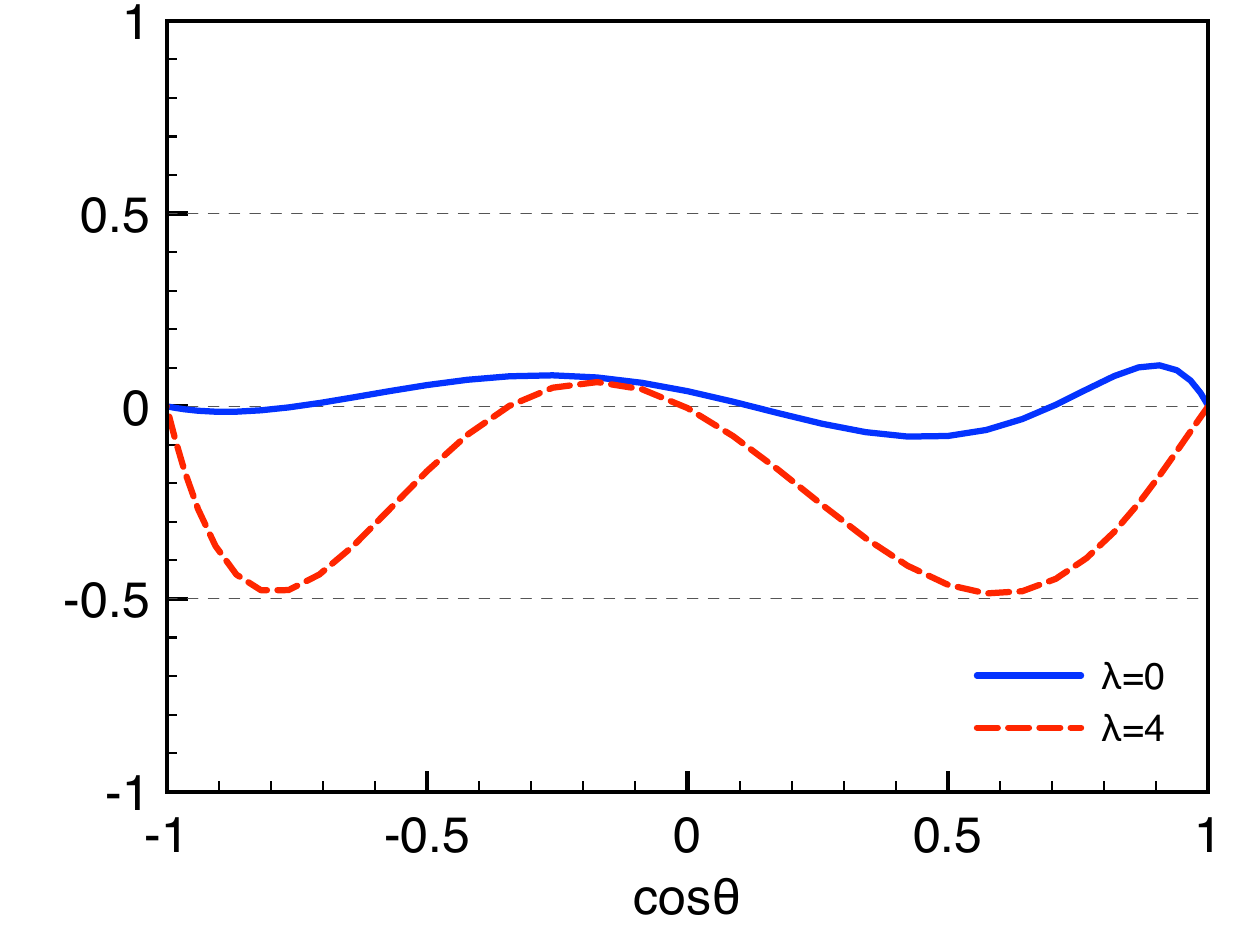}}{-0.4cm}{0.5cm}
\end{tabular}
\caption{(Color online) Spin-density matrix elements (SDMEs) $\rho^\lambda_{00,10,1-1}$ as functions of $\cos\theta$ for the Adair, helicity, and Gottfried-Jackson (GJ) frames for $\lambda=(0,4)$, which stands for the $\phi_4$ helicity ($\pm1$,0), at $W=2.2$ GeV.}
\label{FIG5}
\end{center}
\end{figure}
Fig.~\ref{FIG5} presents spin-density matrix elements (SDMEs) $\rho^{\lambda}_{00,10,1-1}$ in three frames. The $\Delta\lambda_{34}=0$ components dominate at $\cos\theta = \pm1$, indicating helicity conservation. Small violations exist in $f_2$ and $N^*$. 

\begin{figure}[h]
\begin{center}
\begin{tabular}{cc}
\topinset{(a)}{\includegraphics[width= 3.5cm]{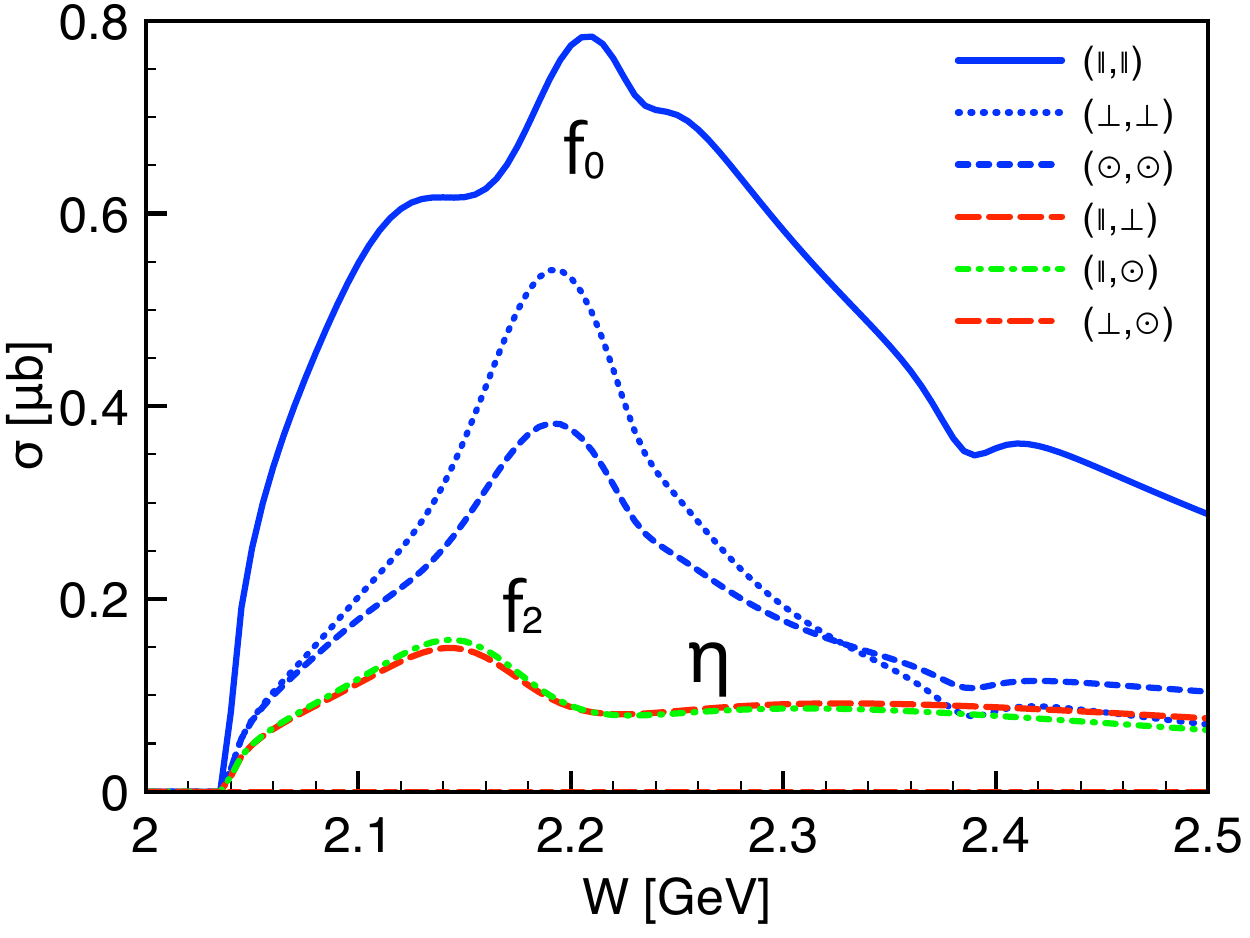}}{-0.3cm}{0.5cm}
\topinset{(b)}{\includegraphics[width= 3.5cm]{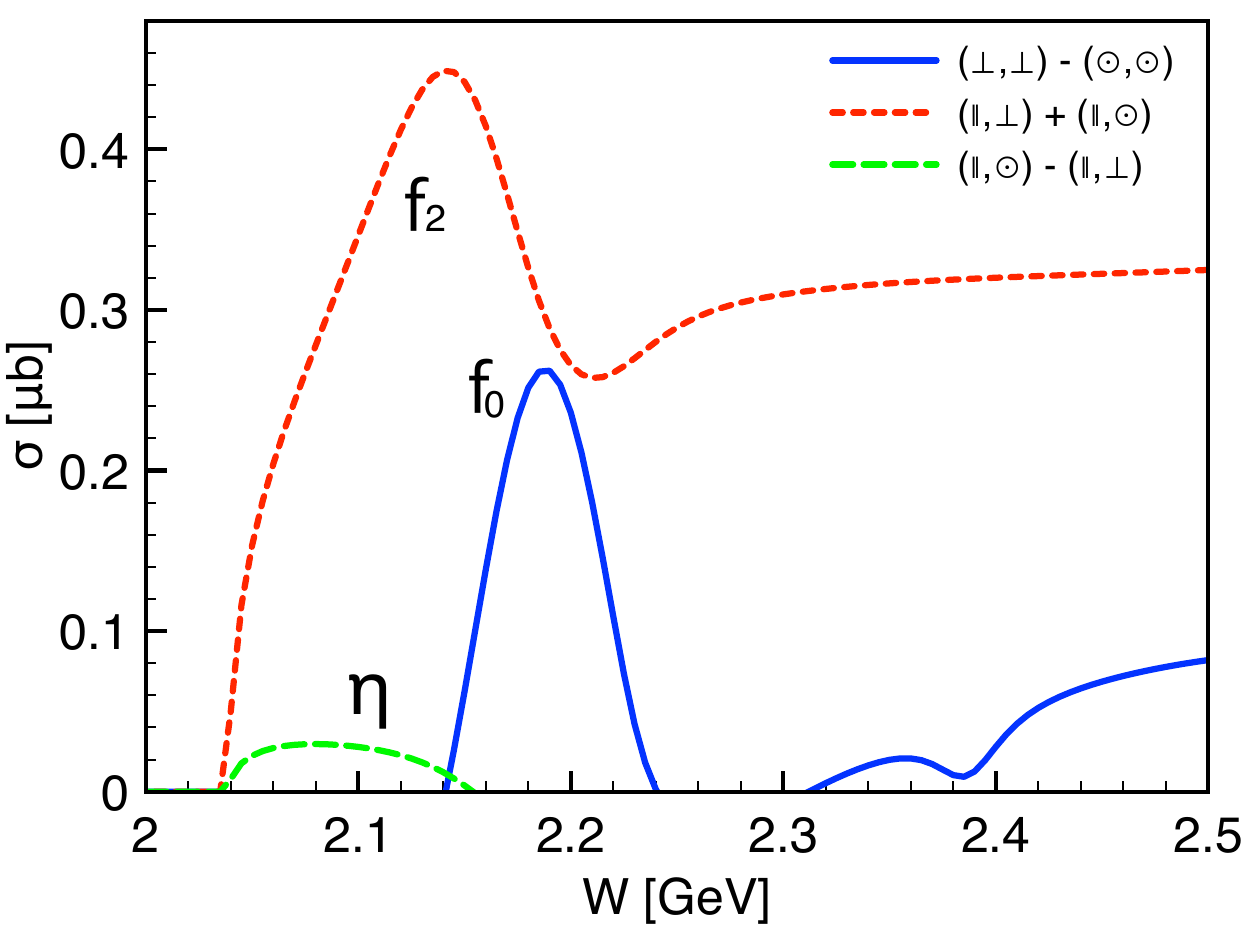}}{-0.3cm}{0.5cm}
\topinset{(c)}{\includegraphics[width= 3.5cm]{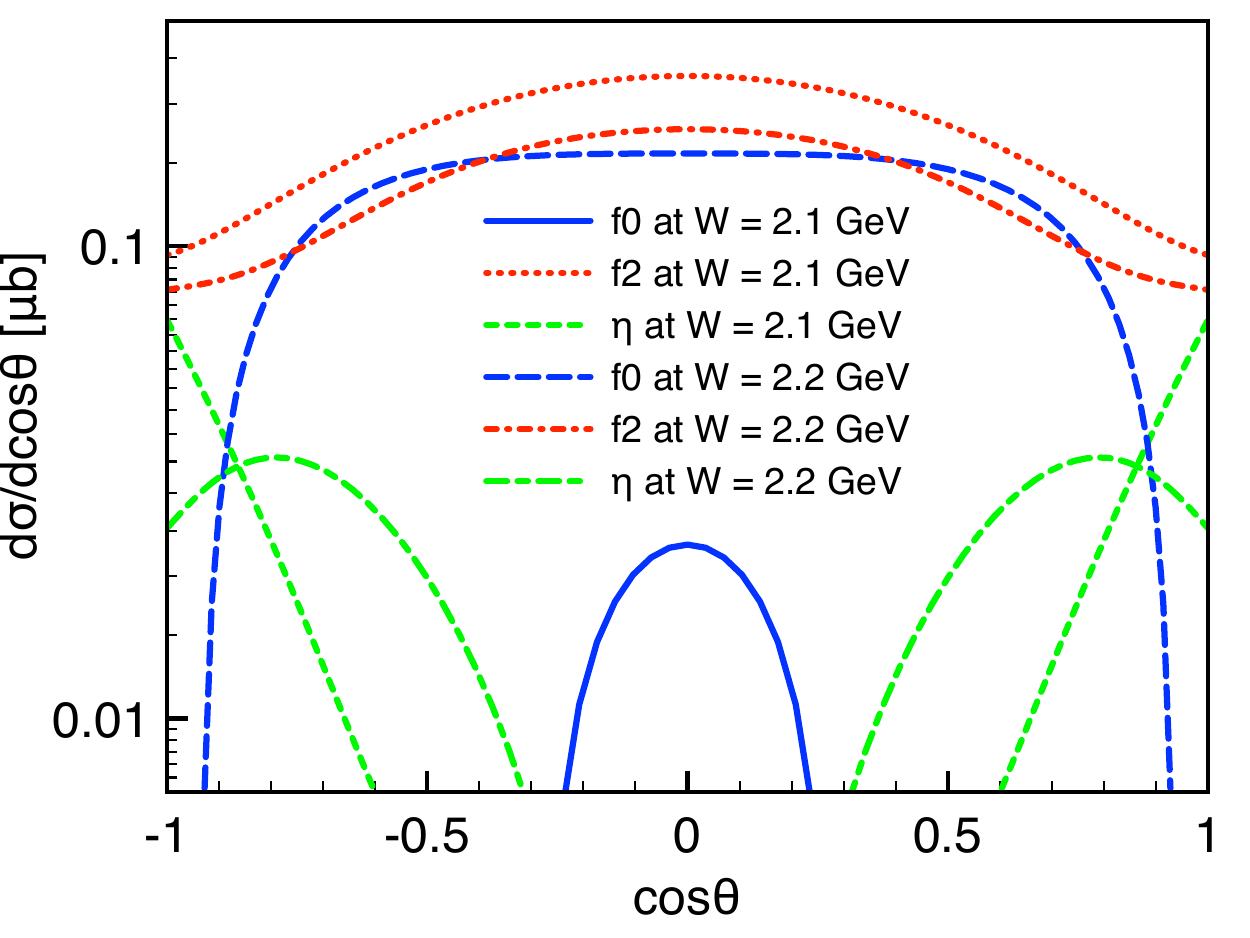}}{-0.3cm}{0.5cm}
\topinset{(d)}{\includegraphics[width= 3.5cm]{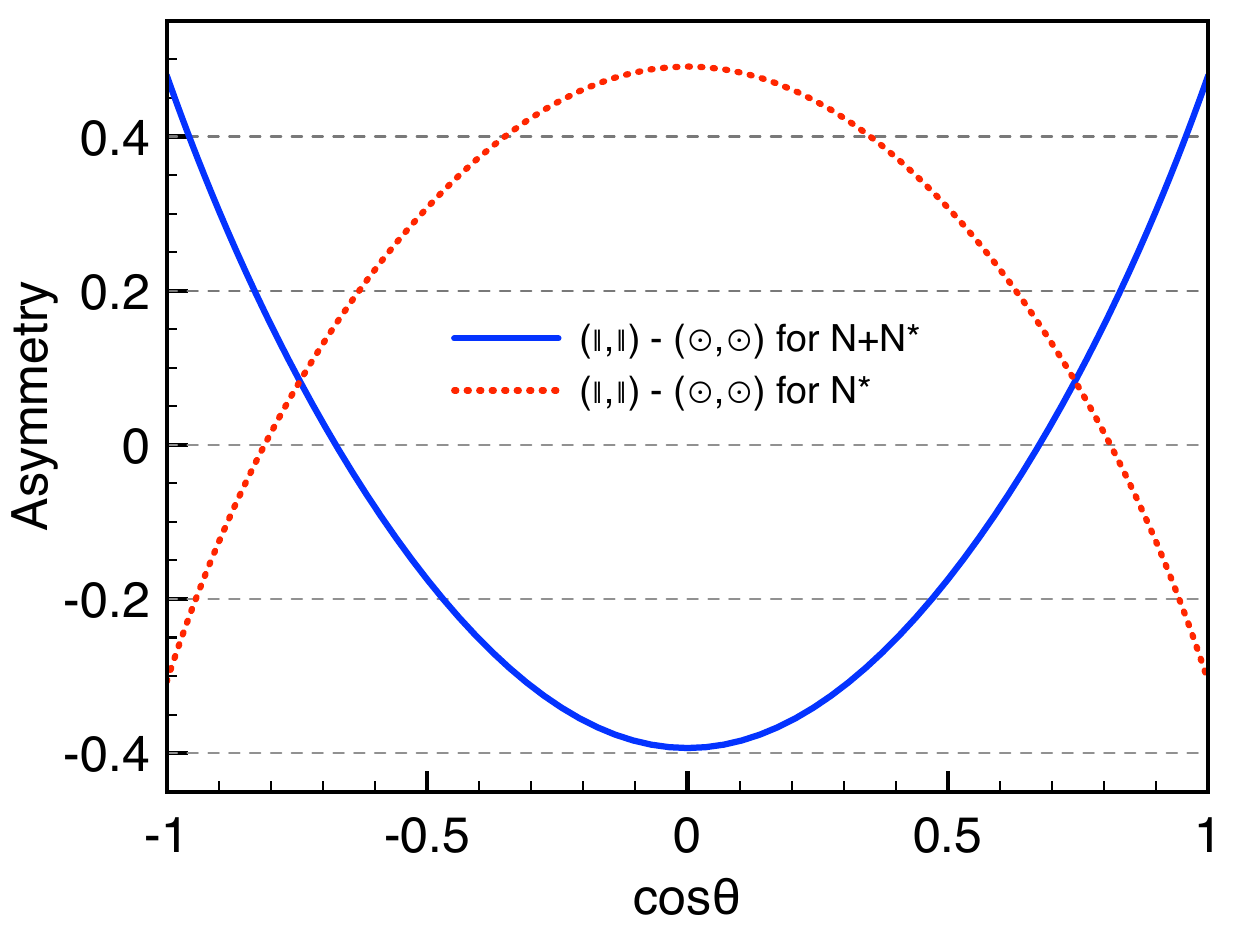}}{-0.3cm}{0.5cm}
\end{tabular}
\caption{(Color online) (a) Polarized total cross-sections as functions of $W$ for the different combinations of $\phi_3$ and $\phi_4$ polarizations, i.e., $(\epsilon_{\phi_3},\epsilon_{\phi_4})$. (b) Added and subtracted polarized cross-sections for the different polarization combinations to enhance the meson signals. (c) Polarized differential cross-sections as functions of $\cos\theta$ in the same manner as the panel (b) for $W=(2.1-2.3)$ GeV. (d) Polarization asymmetries as functions of $\cos\theta$ with $N+N^*$ and $N^*$.}
\label{FIG7}
\end{center}
\end{figure}
Finally, Fig.~\ref{FIG7} shows polarization observables. Cross-sections vary significantly with $(\epsilon_{\phi_3},\epsilon_{\phi_4})$ combinations. By adding or subtracting polarized components, signals from $f_0$, $f_2$, and $\eta$ are enhanced. Panel (d) shows polarization asymmetry differences between $N+N^*$ and $N^*$-only scenarios, suggesting experimental observables that can distinguish between models. More detailed explanations for the numerical calculations can be found in Ref.~\cite{Lee:2024opb}.
\section{Summary and future perspectives}
We investigated double $\phi$ meson production in $\bar{p}p$ annihilation near threshold, focusing on the apparent OZI rule violation. Our effective Lagrangian framework, involving $s$-, $t$-, and $u$-channel exchanges of scalar, tensor, and pseudoscalar mesons, as well as nucleon resonances including a pentaquark-like $P_s$, successfully describes experimental data without invoking explicit quark or gluon degrees of freedom. The total cross-section is primarily shaped by $N$ and $N^*$ contributions near threshold, with distinct $f_0$ and $f_2$ peaks around 2.2 GeV and visible cusp effects near $\bar{\Lambda}\Lambda$ and $\bar{\Sigma}\Sigma$ thresholds. Angular distributions show symmetry at $\theta=\pi/2$, and SDME analysis confirms helicity-conserving dominance. Our findings underscore the significance of hadronic interactions in elucidating OZI-violating processes. 

Future polarization measurements, such as those at J-PARC E104, will be essential to test these predictions. Further theoretical studies, including coupled-channel approaches, are in progress and will be reported elsewhere.
\section*{Acknowledgment}
The authors are grateful for the insightful discussions with Sang-Ho Kim (Soongsil University),  Kanchan Pradeepkumar Khemchandani (Federal University of São Paulo), and Alberto Martinez Torres (University of São Paulo). This work was supported by grants from the National Research Foundation of Korea (NRF), funded by the Korean government (MSIT) (NRF-2022R1A2C1003964, RS-2021-NR060129, and RS-2024-00436392).  The work of D.Y.L. was also supported partially by Global-Learning \& Academic research institution for Master’s, PhD students, and Postdocs (LAMP) Program of the National Research Foundation of Korea(NRF) grant funded by the Ministry of Education (RS-2023-00301702).

\end{document}